\documentclass[12pt]{article}
\usepackage{amsfonts,amssymb,bm,cite,amsmath}
\newcommand {\beq}{\begin{equation}}
\newcommand {\eeq}{\end{equation}}
\newcommand {\beqa}{\begin{eqnarray}}
\newcommand {\eeqa}{\end{eqnarray}}

\allowdisplaybreaks

\begin{document}

\setlength{\oddsidemargin}{0cm}
\setlength{\baselineskip}{7mm}

\begin{titlepage}
\renewcommand{\thefootnote}{\fnsymbol{footnote}}

~~\\

\vspace*{0cm}
    \begin{Large}
       \begin{center}
         {Higher derivative extension of the functional renormalization group }
       \end{center}
    \end{Large}
\vspace{1cm}

\begin{center}
           Gota T{\sc anaka}\footnote
            {
e-mail address : 
tanaka.gota.14@cii.shizuoka.ac.jp}
           and
           Asato T{\sc suchiya}\footnote
           {
e-mail address : 
tsuchiya.asato@shizuoka.ac.jp}\\
      \vspace{1cm}

 {\it Department of Physics, Shizuoka University}\\
                {\it 836 Ohya, Suruga-ku, Shizuoka 422-8529, Japan}\\
         \vspace{0.3cm}     
  {\it Graduate School of Science and Technology, Shizuoka University}\\
                {\it 836 Ohya, Suruga-ku, Shizuoka 422-8529, Japan}
\end{center}

\vspace{3cm}

\begin{abstract}
\noindent
We study higher derivative extension of 
the functional renormalization group (FRG).
We consider FRG equations for a scalar field that consist of terms
with higher functional derivatives of the effective action 
and arbitrary cutoff functions.
We show that the $\epsilon$ expansion around the Wilson-Fisher fixed point 
is indeed reproduced
by the local potential approximation of the FRG equations.
\end{abstract}
\vfill
\end{titlepage}
\vfil\eject

\setcounter{footnote}{0}

\section{Introduction}
\setcounter{equation}{0}
\renewcommand{\thefootnote}{\arabic{footnote}}

The functional renormalization group (FRG) (or the exact renormalization group)
has been developed based on the philosophy given in 
\cite{Wilson:1973jj} (see also \cite{Wegner:1972ih}, and for reviews of the FRG,
see \cite{Morris:1998da,Aoki:2000wm,Bagnuls:2000ae,Polonyi:2001se,Gies:2006wv,Pawlowski:2005xe,Igarashi:2009tj,Rosten:2010vm,Dupuis:2020fhh}).  It serves 
as a powerful method for the nonperturbative study of quantum field theories 
as well as lattice field theories.
The FRG equation is a functional differential equation 
that describes
the dependence of the effective action on the energy scale.

The FRG consists of two procedures, coarse graining and rescaling.
The form of the FRG equation depends on coarse graining procedure.
Thus, one of the important issues on the FRG is 
what coarse graining procedure is allowed.
This issue has been examined thoroughly in the FRG equation 
for a scalar field that includes 
up to two functional derivatives of the effective
action \cite{Latorre:2000qc,Arnone:2002yh,Arnone:2002ai,Arnone:2003pa},
where the coarse graining procedure is fixed by specifying a cutoff function
and a seed action.
It seems, however, 
that the coarse graining procedure that gives the FRG equations 
including more than two functional derivatives 
is quite different from the one that gives the FRG equation including
up to two derivatives.
To our knowledge, such higher derivative FRG equations has not been
studied systematically so far, although there is a recent interesting proposal 
for a manifestly gauge-invariant FRG equation that includes
higher functional derivatives \cite{Sonoda:2020vut} (for related works, see
\cite{Miyakawa:2021hcx,Miyakawa:2021wus,Abe:2022smm,Sonoda:2022fmk,Miyakawa:2022qbz,Makino:2018rys,Abe:2018zdc,Carosso:2018rep,Carosso:2019tan,Sonoda:2019ibh,Matsumoto:2020lha}).

In this paper, we study the FRG equations for a scalar field that 
consist of terms with four (six) functional
derivatives and arbitrary cutoff functions.
We analyze it by using the local potential approximation and
show that the $\epsilon$ expansion around 
the Wilson-Fisher fixed point is 
reproduced if the cutoff functions satisfy certain conditions.
Our result suggests that the coarse graining procedure that gives higher
derivative FRG equations indeed works.

This paper is organized as follows.
In section 2, we briefly review a formal derivation of the FRG equation.
In section 3, we introduce the FRG equations that consist of terms
with four functional derivatives and arbitrary cutoff functions.
In section 4, we analyze the above FRG equations
 by using the local potential
approximation and show that the $\epsilon$ expansion around 
the Wilson-Fisher fixed point is reproduced.
Section 5 is devoted to conclusion and discussion. 
In appendix, we examine the FRG equations consisting of terms with
up to six functional derivatives.

\section{A formal derivation of the FRG equation}
\setcounter{equation}{0}
We consider a scalar field theory in $d$-dimensional Euclidean space.
Throughout this paper,  we use the following notation:
\begin{align}
& \int_x  \ \equiv \int d^d x  \ . 
\end{align}
We denote the effective action at the cutoff scale $\Lambda$ by
$S_{\Lambda}$, which is a functional of the scalar field $\phi(x)$.

The FRG equation follows from an equation \cite{Latorre:2000qc,Arnone:2002yh,Arnone:2005fb,Morris:1999px}
\begin{align}
-\Lambda \frac{\partial}{\partial \Lambda} e^{-S_{\Lambda}}
= \int_x \frac{\delta}{\delta \phi(x)} \left( \Psi_{\Lambda}(x) e^{-S_{\Lambda}} \right)
\label{requirement}
\end{align}
which ensures at least formally
that the partition function
\begin{align}
Z = \int \mathcal{D} \phi \ e^{-S_{\Lambda}}
\end{align}
is independent of $\Lambda$.
Here we emphasize that this derivation of the FRG equation is formal and that one should
check that the equation that follows from (\ref{requirement}) works 
in a physically valid manner.
A standard choice of  $\Psi(x)$ is given by
\begin{align}
\Psi_{\Lambda}(x) = \frac{1}{2}\int_y \dot{C}_{\Lambda}(x-y) \frac{\delta \Sigma_{\Lambda}}{\delta \phi(y)}
\label{Psi}
\end{align}
with 
\begin{align}
\Sigma_{\Lambda}= S_{\Lambda} - 2 \hat{S}_{\Lambda} \ ,
\end{align}
where $C_{\Lambda}(x-y)$ is a cutoff function,
the dot stands for $-\Lambda \frac{d}{d\Lambda}$,
and $\hat{S}_{\Lambda}$ is called the seed action.
The Fourier transform of $\dot{C}_{\Lambda}$, which is defined by
$\dot{C}_{\Lambda}(p)=
\int_x \dot{C}_{\Lambda}(x)e^{-ipx}$, must damp rapidly for
 $|p| > \Lambda$ and allow the Taylor expansion in $p$.

Substituting (\ref{Psi}) into (\ref{requirement}) yields
a class of  FRG equations with up to two functional derivatives:
\begin{align}
-\Lambda \partial_{\Lambda}S_{\Lambda}
=\frac{1}{2}\int_{x,y}\dot{C}_{\Lambda}(x-y) \left(\frac{\delta S_{\Lambda}}{\delta \phi(x)}
\frac{\delta \Sigma_{\Lambda}}{\delta\phi(y)}
-\frac{\delta^2 \Sigma_{\Lambda}}{\delta \phi(x)\delta\phi(y)} \right)
\label{two derivative FRG equation}
\end{align}
The ERG equations (\ref{two derivative FRG equation}) 
are rather general in the sense that they include
arbitrary functions $C_{\Lambda}$ and functionals $\hat{S}_{\Lambda}$.
In particular, putting in (\ref{two derivative FRG equation}) 
\begin{align}
S_{\Lambda}&= \frac{1}{2}\int_{x,y}\phi(x)C_{\Lambda}^{-1}(x-y)\phi(y) + S_I \ ,
\nonumber \\
\hat{S}_{\Lambda}&=\frac{1}{2}\int_{x,y}
\phi(x)C_{\Lambda}^{-1}(x-y)\phi(y)  \ ,
\end{align}
where $S_I$ is the interaction part of the effective action, 
leads to the Polchinski equation \cite{Polchinski:1983gv}
\begin{align}
-\Lambda\partial_{\Lambda}S_I
=\frac{1}{2} \int_{x,y} \dot{C}_{\Lambda}(x-y) 
\left(\frac{\delta S_I}{\delta \phi(x)}
\frac{\delta S_I}{\delta\phi(y)}
-\frac{\delta^2 S_I}{\delta \phi(x)\delta\phi(y)} \right)  \ .
\end{align}
A typical example of $C_{\Lambda}(p)$ is 
\begin{align}
C_{\Lambda}(p)=\frac{e^{-p^2/\Lambda^2}}{p^2} \ .
\end{align}

It was shown in \cite{Latorre:2000qc,Arnone:2002yh,Arnone:2002ai,Arnone:2003pa} that the physical consequences are independent
of the choices of $C_{\Lambda}$ and $\hat{S}_{\Lambda}$.

\section{Higher derivative extension}
\setcounter{equation}{0}
In this section, we consider the FRG equations with four functional derivatives
as a higher derivative extension.
In what follows, we put $t=1/\Lambda^2$, 
which implies that $t\partial_t = -\frac{1}{2}\Lambda 
\partial_{\Lambda}$. We denote $S_{\Lambda}$ by $S_t$.
We consider an equation that follows from (\ref{requirement}).
In order for (\ref{requirement}) to give an FRG equation 
that consists terms with four functional derivatives, $\Psi_{\Lambda}$ must 
consist of terms with three functional derivatives.
All possible types of three functional derivatives are
\begin{align}
\frac{\delta S_t}{\delta \phi(y_1)} \frac{\delta S_t}{\delta \phi(y_2)} \frac{\delta S_t}{\delta \phi(y_3)},  \;\;
\frac{\delta^2 S_t}{\delta \phi(y_1) \delta \phi(y_2)} \frac{\delta S_t}{\delta \phi(y_3)},
\;\;
\frac{\delta^3 S_t}{\delta \phi(y_1) \delta \phi(y_2) \delta \phi(y_3)}  \ .
\end{align}
Thus, we consider almost the most general FRG equation consisting of terms with
four functional derivatives as follows:
\begin{align}
	t \partial_t e^{-S_t} &= \int_x \frac{\delta }{\delta \phi(x)} \left[ \int_{y_1,y_2,y_3} \left\{ A_t(x-y_1)A_t(x-y_2)A_t(x-y_3)
		\frac{\delta S_t}{\delta \phi(y_1)} \frac{\delta S_t}{\delta \phi(y_2)} \frac{\delta S_t}{\delta \phi(y_3)} \right. \right.	\nonumber \\
		&\hspace{100pt} + B_t(x-y_1) B_t(x-y_2) B_t(x-y_3) \frac{\delta^2 S_t}{\delta \phi(y_1) \delta \phi(y_2)} \frac{\delta S_t}{\delta \phi(y_3)} \nonumber	\\
		&\hspace{100pt} \left. \left. + C_t(x-y_1) C_t(x-y_2) C_t(x-y_3) \frac{\delta^3 S_t}{\delta \phi(y_1) \delta \phi(y_2) \delta \phi(y_3)} \right\} e^{-S_t} \right] \ ,	\label{diff4ergeq}
\end{align}
where $A_t(x-y),B_t(x-y)$ and $C_t(x-y)$ are cutoff functions with
the mass dimension $\frac{2}{3}(d-2)$ and assumed to have the following derivative expansions:
\begin{align}
	K_t(x-y) = (K_0 + K_1 \partial_y^2 + \cdots) \delta(x-y)  \ , 
\end{align}
where $K_t$ is $A_t, B_t$ or $C_t$. An example of $K_t$ is 
\begin{align}
	K_t (x-y) &= t^{-\frac{d-2}{3}} t^{\frac{d}{2}}(4 \pi t)^{-\frac{d}{2}} e^{-\frac{(x-y)^2}{4t}}	 \nonumber\\
		&= t^{\frac{d+4}{6}} \int_p e^{-tp^2} e^{ip(x-y)}	\nonumber	\\
		&\simeq t^{\frac{d+4}{6}} \int_p (1-tp^2) e^{ip(x-y)}	\nonumber	\\
		&= t^{\frac{d+4}{6}} \int_p (1+t\partial_x^2) e^{ip(x-y)}	\nonumber	\\
		&= t^{\frac{d+4}{6}} (1+ t \partial_x^2) \delta(x-y)	\ .
\end{align}

\section{Local potential approximation}
\setcounter{equation}{0}
In this section, as a validity check of (\ref{diff4ergeq}),
we analyze it by using the local potential approximation and
show that the $\epsilon$ expansion around 
the Wilson-Fisher fixed point is 
reproduced if the cutoff functions satisfy certain conditions.
\subsection{Flow equation for the local potential}
We apply the local potential approximation \cite{Nicoll:1974zz} to 
(\ref{diff4ergeq}).
First, we represent the effective action in terms of the local potential $V_t$ as
\begin{align}
	S_t[\phi] = \int_x \left(\frac{1}{2}(\partial_x \phi(x)^2) +V_t[\phi](x)\right)  \ .
\end{align}
By substituting this into (\ref{diff4ergeq}), we obtain a flow equation for
the local potential local $V_t[\phi](x)$.
To calculate the first term in the RHS of (\ref{diff4ergeq}), for instance,
we first do the following preparatory calculation:
\begin{align}
	\int_y A_t(x-y) \frac{\delta S_t}{\delta \phi(y)} &=  \int_y A_t(x-y) \frac{\delta}{\delta \phi(y)} \int_{y'} \left[ V_t(y') + \frac{1}{2} (\partial_{y'} \phi(y'))^2 \right]	\nonumber	\\
	&=  \int_{y,y'} A_t(x-y) \left( V'_t(y') - \partial_{y'}^2 \phi(y') \right) \delta(y-y')	\nonumber	\\
	&=  \int_y A_t(x-y) \left( V'_t(y) - \partial_y^2 \phi(y) \right)	\nonumber	\\
	&\simeq  \int_y \left\{ (A_0 + A_1 \partial_y^2) \delta(x-y) \right\} \left( V'_t(y) - \partial_y^2 \phi(y) \right)	\nonumber	\\
	&=  \int_y (A_0 + A_1 \partial_y^2) \left( V'_t(y) - \partial_y^2 \phi(y) \right) \delta(x-y)	\nonumber	\\
	&\simeq  A_0 V_t'(x)   \ ,
\end{align}
\begin{align}
	\int_y A_t(x-y) \frac{\delta^2 S_t}{\delta \phi(x) \delta \phi(y)} &=  \int_y A_t(x-y) \frac{\delta}{\delta \phi(x)} \left( V'_t(y) - \partial_y^2 \phi(y) \right)	\nonumber	\\
	&=  \int_y \frac{\delta}{\delta \phi(x)} \left\{ A_t(x-y) V'_t(y) - \phi(y) \partial_y^2 A_t(x-y) \right\}	\nonumber	\\
	&=  \int_y \left\{ A_t(x-y) V''_t(y) - \partial_y^2 A_t(x-y) \right\} \delta(x-y)	\nonumber	\\
	&=  A_t(0) V''_t(x) + A''_t(0) \ ,
\end{align}
where we used the derivative expansion of the cutoff function $A_t(x-y)$,
$A_t(x-y) \sim (A_0 + A_1 \partial_y^2) \delta(x-y)$.
Using this result, we can calculate the first term  in the RHS of (\ref{diff4ergeq}) as
\begin{align}
	&\int_{x,y,z,w} \frac{\delta}{\delta \phi(x)} \left\{ A_t(x-y) A_t(x-w) A_t(x-z) \frac{\delta S_t}{ \delta \phi(y)} \frac{\delta S_t}{ \delta \phi(z)} \frac{\delta S_t}{ \delta \phi(w)} e^{-S_t} \right\}	\nonumber	\\
	&= \int_{x,y,z,w} A_t(x-y) A_t(x-w) A_t(x-z) \left[ 3 \frac{\delta^2 S_t}{ \delta \phi(x) \delta \phi(y)} \frac{\delta S_t}{ \delta \phi(z)} \frac{\delta S_t}{ \delta \phi(w)} - \frac{\delta S_t}{ \delta \phi(x) }\frac{\delta S_t}{ \delta \phi(y)} \frac{\delta S_t}{ \delta \phi(z)} \frac{\delta S_t}{ \delta \phi(w)} \right]	 e^{-S_t}\nonumber	\\
	&\simeq \int_x \biggl[3 \left\{ - A_t(0) V''_t(x) + A''_t(0) \right\} (- A_0 V_t'(x))^2 + (- A_0 V_t'(x))^3 V_t'(x) \biggr] e^{-S_t}
\end{align}
We can calculate the second and third terms in a similar way and
finally obtain the flow equation for the local potential $V_t$:
\begin{align}
	\partial_t V_t[\phi](x) = PV'' + QV'^2 + LV''V'^2 + MV'^4+XV''^2+YV''''+ZV'''V' \ ,
\label{flow eq for V}
\end{align}
where the coefficients $P,\ Q,\ L,\ M,\ X, \ Y$ and $Z$ are represented as
\begin{align}
	P &= B_{10} B''(0) + B_{20}B(0)	\	,	\nonumber	\\
	Q &= 3A''(0) A_0^2 - B_{20} B_0	\	,	\nonumber	\\
	X &= -B_{10} B(0)	\	,	\nonumber	\\
	Y &= - C^3(0)	\	,	\nonumber	\\
	Z &= -B^2(0) B_0 + C_{10}	\	,		\nonumber	\\
	L &= - 3A(0) A_0^2 + B_{10} B_0	\	,	\nonumber	\\
	M &= A_0^3	\	,	\label{coefficients}
\end{align}
using the derivative expansions of $A_t(x-y), \ B_t(x-y)$ and  $C_t(x-y)$,
\begin{align}	
	A_t(x-y) &\sim (A_0 + A_1 \partial_y^2) \delta(x-y) \ ,	\nonumber\\
	B_t(x-y) &\sim (B_0 + B_1 \partial_y^2) \delta(x-y) \ ,	\nonumber\\
	B^2_t(x-y) &\sim (B_{10}+B_{11} \partial_y^2) \delta(x-y) \ ,	\nonumber\\
	B_t(x-y) \partial_y^2 B_t(x-y) &\sim (B_{20}+B_{21} \partial_y^2) \delta(x-y) \ ,
\nonumber\\
	C^3_t(x-y) &\sim (C_{10} + C_{10} \partial_y^2) \delta(x-y) \ .
\end{align}

Next, we rewrite (\ref{flow eq for V}) in terms of dimensionless quantities.
Note that this procedure realizes the rescaling in the renormalization group.
We add the bar to the dimensionless quantities.
The field $\phi(x)$ and the local potential $V_t$ are made dimensionless as
\begin{align}
    \phi(x) &= t^{-\frac{d-2}{4}} \bar{\phi}(\bar{x})   \ ,  \nonumber\\
	V_t[\phi] &= \bar{V}_t[\bar{\phi}] t^{-\frac{d}{2}} \ ,
\end{align}
respectively.
Thus, the LHS of (\ref{flow eq for V}) is calculated as
\begin{align}
	t\partial_t \bar{V}_t[\phi] &= t \partial_t (\bar{V}_t[\bar{\phi}] t^{-\frac{d}{2}})	\nonumber	\\
	&= t^{-\frac{d}{2}} \left\{ -\frac{d}{2} \bar{V}_t + \frac{d-2}{4} \bar{\phi} \bar{V}_t +t \partial_t \bar{V}_t \right\} \ ,
\end{align}
while the RHS of (\ref{flow eq for V}) as
\begin{align}
	&PV_t'' + QV_t'^2 + LV_t''V_t'^2 + MV_t'^4+XV_t''^2+YV_t''''+ZV_t'''V_t'	\nonumber	\\ 
	&= t^{-\frac{d}{2}} (\bar{P}\bar{V}_t'' + \bar{Q}\bar{V}_t'^2 + \bar{L}\bar{V}_t''\bar{V}_t'^2 +\bar{M}\bar{V}_t'^4+\bar{X}\bar{V}_t''^2+\bar{Y}\bar{V}_t''''+\bar{Z}_t\bar{V}_t'''\bar{V}_t')  \ .
\end{align}
The resultant flow equation for $\bar{V}_t$ is 
\begin{align}
	t \partial_t \bar{V}_t = \frac{d}{2} \bar{V}_t - \frac{d-2}{4} \bar{\phi} \bar{V}_t + \bar{P}\bar{V}_t'' + \bar{Q}\bar{V}_t'^2 + \bar{L}\bar{V}_t''\bar{V}_t'^2 +\bar{M}\bar{V}_t'^4+\bar{X}\bar{V}_t''^2+\bar{Y}\bar{V}_t''''+\bar{Z}\bar{V}_t'''\bar{V}_t'  \ .	\label{dimensionless flow eq for V}
\end{align}
In what follows, we omit the bar for dimensionless quantities.

Finally, we expand the local potential$V_t$ in $\phi$ to the eighth order as
\begin{align}
	V_t[\phi](x) = \frac{v_2}{2!} \phi^2(x) + \frac{v_4}{4!} \phi^4(x) + \frac{v_6}{6!} \phi^6(x) + \frac{v_8}{8!} \phi^8(x) \ ,
\label{expansion of V}
\end{align}
where the $Z_2$ symmetry is assumed,
and substitute (\ref{expansion of V}) into (\ref{dimensionless flow eq for V}).
This results in the following flow equations for the couplings $v_{2n}$:
\begin{align}
	t \partial_t v_2 &= 2 L v_2^3+P v_4+2 Q v_2^2+2 v_2 v_4 (X+Z)+v_2+v_6 Y	\,	\nonumber	\\
	t \partial_t v_4 &= 20 L v_2^2 v_4+24 M v_2^4+P v_6+8 Q v_2 v_4+2 v_2 v_6 X+4 v_2 v_6 Z+6 v_4^2 X+4 v_4^2 Z+\frac{v_4 \epsilon }{2}+v_8 Y	\,	\nonumber		\\
	t \partial_t v_6 &= 42 L v_2^2 v_6+140 L v_2 v_4^2+480 M v_2^3 v_4+P v_8+4 Q \left(3 v_2 v_6+5 v_4^2\right)+2 v_2 v_8 X	\nonumber	\\
		& \;\;\; +6 v_2 v_8 Z+30 v_4 v_6 X+26 v_4 v_6 Z+v_6 \epsilon -v_6	\,	\nonumber		\\
	t \partial_t v_8 &= 8 L \left(9 v_2^2 v_8+126 v_2 v_4 v_6+70 v_4^3\right)+1344 M v_2^2 \left(v_2 v_6+5 v_4^2\right)+16 Q v_2 v_8	\nonumber	\\
		& \;\;\; +112 Q v_4 v_6+56 v_4 v_8 X+64 v_4 v_8 Z+70 v_6^2 X+56 v_6^2 Z+\frac{3 v_8 \epsilon }{2}-2 v_8	\,	\label{flow equations for couplings}
\end{align}
where we put $d=4-\epsilon$.

\subsection{Fixed points}
The fixed points of the renormalization group are determined by
\begin{align}
\partial_t v_{2n} = 0 \ .
\label{fixed point equation}
\end{align}
in (\ref{flow equations for couplings}).
We denote a solution to (\ref{fixed point equation}) by 
$v_{2n}^*$.

We perform the $\epsilon$ expansion to the first order of 
$\epsilon$ in the following (for the $\epsilon$ expansion for
the FRG consisting of up to two functional derivatives, see 
\cite{ODwyer:2007brp}).
We find a trivial fixed point, the Gaussian fixed point  given by
\begin{align}
	v_2^* &= 0,\hspace{10pt} v_4^* = 0, \hspace{10pt} v_6^* = 0,\hspace{10pt} v_8^* = 0 \ ,
\end{align}
and a nontrivial fixed point, the Wilson-Fisher
fixed point given by
\begin{align}
	v_2^* &= \frac{P}{4 (6 P Q+3 X+2 Z)} \epsilon	+ \mathcal{O}(\epsilon^2)  \ ,	\nonumber	\\
	v_4^* &= -\frac{1}{4 (6 P Q+3 X+2 Z)}	 \epsilon + \mathcal{O}(\epsilon^2)  \ ,	\nonumber \\
	v_6^* &= \frac{5 Q}{4 (6 P Q+3 X+2 Z)^2} \epsilon^2 + \mathcal{O}(\epsilon^3)  \ ,	\nonumber	\\
	v_8^* &= -\frac{35 \left(L+4 Q^2\right)}{8 (6 P Q+3 X+2 Z)^3} \epsilon^3	 + \mathcal{O}(\epsilon^4)  \ .
	\label{Wilson-Fisher fixed point}
\end{align}
We see that the following condition must be satisfied in order for the 
Wilson-Fisher fixed point to exist:
\begin{align}
	6PQ + 3X + 2Z \neq 0. \ .
\end{align}

Putting $v_{2n} = v_{2n}^* + \delta v_{2n}$,
we linearize the flow equations (\ref{flow equations for couplings}) around 
the nontrivial fixed (\ref{Wilson-Fisher fixed point}) with respect to $\delta v_{2n}$
as follows:
\begin{align}
	\partial_t \delta v = T \delta v  \ ,
\end{align}
where
\begin{align}
\delta v &=  {}^t (\delta v_2, \delta v_4, \delta v_6, \delta v_8)  \ , \\
T&=
\left(
\begin{array}{cccc}
 T_{11} & T_{12} & T_{13} & T_{14} \\
 T_{21} & T_{22} & T_{23} & T_{24} \\
 T_{31} & T_{32} & T_{33} & T_{34} \\
 T_{41} & T_{42} & T_{43} & T_{44} \\
\end{array}
\right).  \ ,	\nonumber\\
	T_{11} &= 1 - \frac{-2 PQ + X + Z}{12 PQ + 6X + 4Z} \epsilon + O(\epsilon^2)	\ ,	\nonumber	\\
	T_{12} &= P + \frac{P (X+Z)}{12 P Q+6 X+4 Z} \epsilon + O\left(\epsilon ^2\right)	\ ,	\nonumber	\\
	T_{13} &= Y	\ ,	\nonumber	\\
	T_{14} &= 0	\ ,	\nonumber	\\
	T_{21} &= -\frac{2 Q }{6 P Q+3 X+2 Z}  \epsilon + O\left(\epsilon ^2\right)	\ ,	\nonumber	\\
	T_{22} &= \frac{10 P Q-3 X-2 Z}{12 P Q+6 X+4 Z} \epsilon + O\left(\epsilon ^2\right)	\ ,	\nonumber	\\
	T_{23} &= P + \frac{P  (X+2 Z)}{12 P Q+6 X+4 Z} \epsilon + O\left(\epsilon ^2\right)	\ ,	\nonumber	\\
	T_{24} &= Y	\ ,	\nonumber	\\
	T_{31} &= O(\epsilon )^2	\ ,	\nonumber	\\
	T_{32} &= -\frac{10 Q }{6 P Q+3 X+2 Z} \epsilon + O\left(\epsilon ^2\right)	\ ,	\nonumber	\\
	T_{33} &= -1 + \frac{ 18 P Q-9 (X+Z)}{12 P Q+6 X+4 Z} \epsilon + O\left(\epsilon ^2\right)	\ ,	\nonumber	\\
	T_{34} &= P + \frac{ P  (X+3 Z)}{12 P Q+6 X+4 Z} \epsilon + O\left(\epsilon ^2\right)	\ ,	\nonumber	\\
	T_{41} &= O\left(\epsilon ^2\right)	\ ,	\nonumber	\\
	T_{42} &= O\left(\epsilon ^2\right)	\ ,	\nonumber	\\
	T_{43} &= -\frac{28 Q }{6 P Q+3 X+2 Z} \epsilon + O\left(\epsilon ^2\right)	\ ,	\nonumber	\\
	T_{44} &= -2 + \frac{  26 P Q-19 X-26 Z }{12 P Q+6 X+4 Z} \epsilon + O\left(\epsilon ^2\right)	\	.
\end{align}

The eigenvalues of $T$ are calculated up to the first order in $\epsilon$  as
\begin{align}
	\lambda_2 &= 1-\frac{2 P Q+X+Z}{12 P Q+6 X+4 Z}\epsilon \ ,	\nonumber\\
	\lambda_4 &= -\frac{1}{2}\epsilon \ ,	  \nonumber\\
	\lambda_6 &= -1 -\frac{9 (2 P Q+X+Z)}{12 P Q+6 X+4 Z}\epsilon \ ,	         
     \nonumber\\
	\lambda_8 &= -2 + \frac{82 P Q-19 X-26 Z}{12 P Q+6 X+4 Z} \epsilon \ .	
\label{eigenvalues}
\end{align}
The eigenvalues $\lambda_2$, $\lambda_4$, $\lambda_6$ 
and $\lambda_8$ are supposed to be fixed by
by the scaling dimension of the operators $\int_x\phi^2,\ \int_x\phi^4,\ \int_x\phi^6$ and $\int_x\phi^8$,
respectively, as
\begin{align}
	\lambda_2 = 1 - \frac{\epsilon}{6}, \;
	\lambda_4 = - \frac{\epsilon}{2}, \; 
	\lambda_6 = -1 - \frac{3 \epsilon}{2}, \;
	\lambda_8 = -2 - \frac{19 \epsilon}{6} \ .
\label{scaling dimensions}
\end{align}
We see that 
$\lambda_2, \ \lambda_4$ and $\lambda_6$ with $Z=0$ in (\ref{eigenvalues})
 indeed agree those in
(\ref{scaling dimensions}).
Note that $\lambda_8$ with $Z=0$ in (\ref{eigenvalues}) does not agrees with that in (\ref{scaling dimensions}).
This is because the local potential is truncated up to the eighth order in $\phi$.
We verified that we
obtain the correct value of $\lambda$ if we expand the local potential
to the tenth order in $\phi$ and performed the same analysis.

As a consequence,
in order that the $\epsilon$ expansion with the local potential approximation
gives the correct values of the scaling dimensions around the Wilson-Fisher
fixed point, the following two conditions must be satisfied:
\begin{align}
	2PQ + X \neq 0 &\Leftrightarrow 2(B''(0) B_{10} + B(0) B_{20})( 3 A''(0) A_0^2 - B_0 B_{20})-B(0) B_{10} \neq 0	\label{criteria1}	\\
	Z = 0 &\Leftrightarrow - B^2(0) B_0 + C_{10}=0	\label{criteria2} 
\end{align}
Namely, the cutoff functions $A_t(x-y), \ B_t(x-y)$ and $C_t(x-y)$
in (\ref{diff4ergeq}) must be chosen such that these two conditions are 
satisfied.

We have analyzed the FRG equation with terms consisting of
four functional derivatives
so far. We can generalize the above analysis to the cases in which
the FRG equations include terms with two or more than four functional
derivatives in addition to the terms with four functional derivatives.
In these cases, we can show that the $\epsilon$ expansion with the
local potential approximation reproduces the scaling dimensions 
to the first order in $\epsilon$ if the conditions (\ref{criteria1}) and
(\ref{criteria2}) are satisfied. In appendix, we examine the local 
potential approximation for the FRG equation including terms
with two, four or six functional derivatives.

Our results suggest that the FRG equation can be extended such that it
includes higher functional derivatives.

\section{Conclusion and discussion}
\setcounter{equation}{0}
In this paper, we studied the higher derivative extension of the FRG.
We considered the FRG equations for a scalar field that consists
of the terms with four functional derivatives
and arbitrary cutoff functions.
While those FRG equations are constructed in such a way that they
guarantee the invariance of the partition function under the changes of
scale at least formally,
it is nontrivial that they make sense physically because the coarse graining
corresponding to four functional derivatives is quite different from that to
two functional derivatives.
We showed that the $\epsilon$ expansion around the Wilson-Fisher fixed point 
is indeed reproduced
by the local potential approximation of the FRG equations
if the cutoff functions satisfy the conditions. We also verified that
this holds for the case of six functional derivatives.
It is natural that the conditions on the cutoff functions are needed
because it is known that the derivative expansion for the FRG equations,
whose lowest order is nothing but the local potential approximation\footnote{ It was shown
in \cite{Rosten:2010vm} that the local potential approximation for the FRG equations consisting
of terms with up to two functional derivatives has no dependence on the cutoff function.},
in general breaks
the arbitrariness of the cutoff functions and the invariance under redefinition
of the field (see \cite{Rosten:2010vm} and references therein.)\footnote{Note also that we did not consider in this paper 
$\Sigma_{\Lambda}$ that includes the functional derivative of $S_t$ and
depends explicitly on $\phi$ such as
\begin{align}
	\Sigma_{\Lambda} = \int_y K_t(x-y) \phi^2(y) \frac{\delta S_t}{\delta \phi(y)} \ .	\nonumber
\end{align}
We saw that this type of $\Sigma$ in the FRG equations for a scalar field prevents the local potential approximation from reproducing
the known scaling dimensions.}. 
Our results suggest that the higher derivative extension of
the FRG makes sense.

\section*{Acknowledgments}
A.T. was supported in part by Grant-in-Aid for Scientific Research (No. 18K03614 and No. 21K03532) from
Japan Society for the Promotion of Science.

\appendix
\section{Local potential approximation for the FRG equations with 
up to six functional derivatives}
Here we examine the local potential approximation for the FRG equation
including terms with two, four or six functional derivatives.
In this case, the local approximation yields the following flow equation
for the local potential $V_t$:
\begin{align}
	t \partial_t V_t &= P_1 V_1^2+P_2 V_2+Q_5 V_1^4+Q_4 V_2 V_1^2+Q_3 V_3 V_1+Q_1 V_2^2+Q_2 V_4+R_{11} V_1^6	\nonumber	\\
		&+R_{10} V_2 V_1^4+R_7 V_3 V_1^3+R_9 V_2^2 V_1^2+R_4 V_4 V_1^2	\nonumber	\\
		&+R_6 V_2 V_3 V_1+R_2 V_5 V_1+R_8 V_2^3+R_5 V_3^2+R_3 V_2 V_4+R_1 V_6. \ ,	\label{ergeqforv6}
\end{align}
where $V_n$ stands for the $n$-th order derivative of $V_t$ with respect to $\phi$,
and $P_i$, $Q_i$ and $R_i$ are determined by the cutoff functions as in (\ref{coefficients}).
Making the above flow equation dimensionless and 
substituting (\ref{expansion of V}) into (\ref{ergeqforv6}) yields
\begin{align}
	t \partial_t v_2 &= 2 P_1 v_2^2+P_2 v_4+2 Q_4 v_2^3+2 Q_1 v_2 v_4+2 Q_3 v_2 v_4+Q_2 v_6+2 R_4 v_2^2 v_4+2 R_6 v_2^2 v_4 	\nonumber	\\
		&+3 R_8 v_2^2 v_4+2 R_9 v_2^4+2 R_2 v_2 v_6+R_3 v_2 v_6+R_3 v_4^2+2 R_5 v_4^2+R_1 v_8+v_2  \ ,	\nonumber\\
	t \partial_t v_4 &= 8 P_1 v_2 v_4+P_2 v_6+20 Q_4 v_2^2 v_4+24 Q_5 v_2^4+2 Q_1 v_2 v_6+4 Q_3 v_2 v_6+6 Q_1 v_4^2+4 Q_3 v_4^2+Q_2 v_8	\nonumber	\\
		&+24 R_7 v_2^3 v_4+32 R_9 v_2^3 v_4+12 R_4 v_2^2 v_6+4 R_6 v_2^2 v_6+3 R_8 v_2^2 v_6+24 R_{10} v_2^5+8 R_4 v_2 v_4^2+16 R_6 v_2 v_4^2	\nonumber	\\
		&+18 R_8 v_2 v_4^2+4 R_2 v_2 v_8+R_3 v_2 v_8+4 R_2 v_4 v_6+7 R_3 v_4 v_6+8 R_5 v_4 v_6+\frac{v_4 \epsilon }{2}	  \ , \nonumber\\
	t \partial_t v_6 &= 4 P_1 \left(3 v_2 v_6+5 v_4^2\right)+P_2 v_8+480 Q_5 v_2^3 v_4+42 Q_4 v_2^2 v_6+140 Q_4 v_2 v_4^2+2 Q_1 v_2 v_8	\nonumber	\\
		&+6 Q_3 v_2 v_8+30 Q_1 v_4 v_6+26 Q_3 v_4 v_6+360 R_7 v_2^2 v_4^2+440 R_9 v_2^2 v_4^2+840 R_{10} v_2^4 v_4+120 R_7 v_2^3 v_6	  
		\nonumber	\\
		&+ 72 R_9 v_2^3 v_6+30 R_4 v_2^2 v_8+6 R_6 v_2^2 v_8+3 R_8 v_2^2 v_8+720 R_{11} v_2^6+132 R_4 v_2 v_4 v_6+116 R_6 v_2 v_4 v_6	\nonumber	\\
		&+90 R_8 v_2 v_4 v_6+20 R_4 v_4^3+60 R_6 v_4^3+90 R_8 v_4^3+20 R_2 v_4 v_8+16 R_3 v_4 v_8+12 R_5 v_4 v_8	\nonumber	\\
		&+6 R_2 v_6^2+15 R_3 v_6^2+20 R_5 v_6^2+v_6 \epsilon -v_6 \ ,	\nonumber\\
	t \partial_t v_8 &= 16 P_1 (v_2 v_8+7 v_4 v_6)+6720 Q_5 v_2^2 v_4^2+1344 Q_5 v_2^3 v_6+72 Q_4 v_2^2 v_8+1008 Q_4 v_2 v_4 v_6	\nonumber	\\
		&+560 Q_4 v_4^3+14 Q_1 \left(4 v_4 v_8+5 v_6^2\right)+64 Q_3 v_4 v_8+56 Q_3 v_6^2+20160 R_{10} v_2^3 v_4^2+4368 R_7 v_2^2 v_4 v_6	 \nonumber	\\
		&+3584 R_9 v_2^2 v_4 v_6+40320 R_{11} v_2^5 v_4+3024 R_{10} v_2^4 v_6+336 R_7 v_2^3 v_8+128 R_9 v_2^3 v_8	\nonumber	\\
		&+3360 R_7 v_2 v_4^3+4480 R_9 v_2 v_4^3+576 R_4 v_2 v_4 v_8+288 R_6 v_2 v_4 v_8+168 R_8 v_2 v_4 v_8	\nonumber	\\
		&+336 R_4 v_2 v_6^2+336 R_6 v_2 v_6^2+210 R_8 v_2 v_6^2+672 R_4 v_4^2 v_6+1008 R_6 v_4^2 v_6+1260 R_8 v_4^2 v_6	\nonumber	\\
		&+64 R_2 v_6 v_8+98 R_3 v_6 v_8+112 R_5 v_6 v_8+\frac{3 v_8 \epsilon }{2}-2 v_8. \ .
\end{align}
The nontrivial fixed point is given by
\begin{align}
	v_2^* &= \frac{P_2}{4 \left(6 P_1 P_2+3 Q_1+2 Q_3\right)} \epsilon \ ,	\nonumber\\
	v_4^* &= -\frac{1}{4 \left(6 P_1 P_2+3 Q_1+2 Q_3\right)} \epsilon	 \ , \nonumber\\
	v_6^* &= \frac{5 P_1}{4 \left(6 P_1 P_2+3 Q_1+2 Q_3\right){}^2} \epsilon^2 \,	\nonumber\\
	v_8^* &= -\frac{35 \left(4 P_1^2+Q_4\right)}{8 \left(6 P_1 P_2+3 Q_1+2 Q_3\right){}^3} \epsilon^3	 \ .
\end{align}
We see that $6 P_1 P_2+3 Q_1+2 Q_3 \neq 0$ is required for the nontrivial
fixed point to exist.
The linearized equation around the nontrivial fixed point is
\begin{align}
	\partial_t \delta v = T \delta v 
\end{align}
with 
\begin{align}
T=
\left(
\begin{array}{cccc}
 T_{11} & T_{12} & T_{13} & T_{14} \\
 T_{21} & T_{22} & T_{23} & T_{24} \\
 T_{31} & T_{32} & T_{33} & T_{34} \\
 T_{41} & T_{42} & T_{43} & T_{44} \\
\end{array}
\right)  \ ,
\end{align}
\begin{align}
	T_{11} &= 1-\frac{\epsilon  \left(-2 P_1 P_2+Q_1+Q_3\right)}{12 P_1 P_2+6 Q_1+4 Q_3}+O\left(\epsilon ^2\right)	   \ , \nonumber\\
	T_{12} &= P_2 +\frac{\epsilon  \left(P_2 \left(Q_1+Q_3\right)-R_3-2 R_5\right)}{12 P_1 P_2+6 Q_1+4 Q_3}	+ O\left(\epsilon ^2\right)	\ , \nonumber\\
	T_{13} &= Q_2+\frac{P_2 \left(2 R_2+R_3\right) \epsilon }{4 \left(6 P_1 P_2+3 Q_1+2 Q_3\right)}+O\left(\epsilon ^2\right)	\ , \nonumber\\
	T_{14} &= R_1 \ , 	\nonumber\\
	T_{21} &= -\frac{2 P_1 \epsilon }{6 P_1 P_2+3 Q_1+2 Q_3} + O\left(\epsilon ^2\right) \ , 	\nonumber\\
	T_{22} &= \frac{\epsilon  \left(10 P_1 P_2-3 Q_1-2 Q_3\right)}{12 P_1 P_2+6 Q_1+4 Q_3} + O\left(\epsilon ^2\right)	\ , \nonumber\\
	T_{23} &= P_2 -\frac{\epsilon  \left(-2 P_2 \left(Q_1+2 Q_3\right)+4 R_2+7 R_3+8 R_5\right)}{4 \left(6 P_1 P_2+3 Q_1+2 Q_3\right)} + O\left(\epsilon ^2\right)	\ ,    \nonumber\\
	T_{24} &= Q_2 + \frac{P_2 \left(4 R_2+R_3\right) \epsilon }{4 \left(6 P_1 P_2+3 Q_1+2 Q_3\right)} + O\left(\epsilon ^2\right)	\ ,  \nonumber\\
	T_{31} &= O(\epsilon )^2	\ , \nonumber\\
	T_{32} &= -\frac{10 P_1 \epsilon }{6 P_1 P_2+3 Q_1+2 Q_3} + O\left(\epsilon ^2\right)	\ ,   \nonumber\\
	T_{33} &= -1+\frac{\epsilon  \left(18 P_1 P_2-9 \left(Q_1+Q_3\right)\right)}{12 P_1 P_2+6 Q_1+4 Q_3}+O\left(\epsilon ^2\right) \ , 	\nonumber\\
	T_{34} &= P_2 + \frac{\epsilon  \left(P_2 \left(Q_1+3 Q_3\right)-2 \left(5 R_2+4 R_3+3 R_5\right)\right)}{12 P_1 P_2+6 Q_1+4 Q_3} + O\left(\epsilon ^2\right)	\ , \nonumber\\
	T_{41} &= O\left(\epsilon ^2\right) \ , 	\nonumber\\
	T_{42} &= O\left(\epsilon ^2\right) \ ,  \nonumber\\
	T_{43} &= -\frac{28 P_1 \epsilon }{6 P_1 P_2+3 Q_1+2 Q_3} + O\left(\epsilon ^2\right)	\ ,  \nonumber\\												
	T_{44} &= -2+\frac{\epsilon  \left(26 P_1 P_2-19 Q_1-26 Q_3\right)}{12 P_1 P_2+6 Q_1+4 Q_3}+O\left(\epsilon ^2\right)	\ .
\end{align}
The eigenvalues of $T$ are
\begin{align}
	\lambda_2 &= -1 -\frac{2 P_1 P_2+Q_1+Q_3}{12 P_1 P_2+6 Q_1+4 Q_3} \epsilon	\ , \nonumber\\
	\lambda_4 &= -\frac{1}{2} \epsilon  \ ,	\nonumber\\
	\lambda_6 &= -1 -\frac{9 \left(2 P_1 P_2+Q_1+Q_3\right)}{12 P_1 P_2+6 Q_1+4 Q_3} \epsilon	\ , \nonumber\\
	\lambda_8 &= -2 + \frac{82 P_1 P_2-19 Q_1-26 Q_3}{12 P_1 P_2+6 Q_1+4 Q_3} \epsilon  \ .
\end{align}
$\lambda_1, \; \lambda_2$ and $\lambda_3$ 
agree with those in (\ref{scaling dimensions}) if $Q_3=0$.
Thus, we see that the coefficients of six derivatives in (\ref{ergeqforv6})
are arbitrary, while those of four derivatives must satisfy (\ref{criteria1}) and (\ref{criteria2}).

\end{document}